# Energetics of vacancy segregation to symmetric tilt grain boundaries in HCP materials

M A Bhatia[1], K N Solanki[1*]

[1]School for Engineering of Matter, Transport, and Energy, Arizona State University, Tempe, AZ 85287, USA
*Corresponding Author: (480) 965-1869; (480) 727-9321 (fax), E-mail: kiran.solanki@asu.edu

## Abstract

Molecular static simulations of 190 symmetric tilt grain boundaries in HCP metals were used to understand the energetics of vacancy segregation, which is important for designing stable interfaces in harsh environments. Simulation results show that the local arrangements of grain boundaries and the resulting structural units have a significant influence on the magnitude of vacancy binding energies, and the site-to-site variation within each boundary is substantial. Comparing the vacancy binding energies for each site in different c/a ratio materials shows that the binding energy increases significantly with an increase in c/a ratio. For example, in the $[1\bar{2}10]$ tilt axis, Ti and Zr with c/a=1.5811 have a lower vacancy binding energy than the Mg with c/a=1.6299. Furthermore, when the grain boundary energies of all 190 boundaries in all three elements are plotted against the vacancy binding energies of the same boundaries, a highly negative correlation ($r$ = -0.7144) is revealed that has a linear fit with a proportionality constant of -25 Å$^2$. This is significant for applications where extreme environmental damage generates lattice defects and grain boundaries act as sinks for both vacancies and interstitial atoms.

## I Introduction

Increasing global demand for safer, energy-efficient, bio-compatible systems for biomedical, transportation, and safety applications requires development of new materials with tuned interface structures.[1] Because the mechanical behavior of polycrystalline material is often driven by grain boundaries and their underlying structure, [2–4] a fundamental understanding of the relationship between the grain boundary structure and associated properties is important to the development of interface-dominant materials. The term *grain boundary character* is often used to describe the five degrees of freedom necessary to define a grain boundary. Three degrees of freedom are used to define the misorientation between the two grains, and two degrees of freedom are associated with the grain boundary plane. Research has shown that both the macroscopic degrees of freedom and microscopic local structure affect the physical properties of grain boundaries.[5–11] In terms of the microscopic local structure, the translations between adjoining grains are also important, as is the localized dislocation structure of the boundary.

Historically, research has focused on developing a method to characterize grain boundaries[12–17] and their influence on the physical properties of polycrystalline materials. These models utilize dislocation arrays, disclinations, and coincident site lattice (CSL) to describe the local structure of grain boundaries. These efforts, in turn, have led to identifying the primary structural elements for symmetric tilt, asymmetric tilt, twist, and twin boundaries at the atomic scale.[5,8,18–24] The term structural unit (SU) has been used to describe the local atomic arrangement at the grain boundary and is associated with both the grain boundary character and its properties (e.g., see

[10,11,25–27]). Saylor et al.[28] experimentally, studied the grain boundary character distribution (GBCD) as a function of grain boundary geometry for a commercially pure aluminum (Al) sample. They indicated that boundaries with lower energies and index planes have a higher distribution in the polycrystalline sample. These results also apply to other metals such as nickel (Ni) and copper (Cu). Grain boundary structure has also been observed using field ion microscopy and high resolution transmission electron microscopy.[29–33] In addition, grain boundary energies can be computed through theoretical formulations and computational methods. The role of the grain boundary plane in determining the grain boundary energy was investigated by Wang and Beyerlein,[34,35] who performed extensive calculations of the symmetric tilt grain boundary (STGB) energies for hexagonal closed pack (HCP) metals. The aforementioned research notwithstanding, the role of grain boundary character on the energetics of point defects' segregation at the interface has received less attention, especially in HCP materials with varying degrees of grain boundary SUs.

Quantifying how point defects interact with defect sinks, such as grain boundaries, is also important for understanding strength of material interfaces in various environments, such as titanium (Ti) in a high oxygen environment, zirconium (Zr) in an irradiation environment, and magnesium (Mg) in a corrosive environment. For instance, during irradiation-induced segregation, the flux of solute and impurity elements is highly coupled with the flux of vacancies and interstitials. As vacancies and interstitials tend to diffuse and bind to microstructural sinks, solute and impurity atoms are spatially redistributed in the vicinity of these sinks.[36,37] The net result is either accumulation or depletion of elements at these defect sinks, both of which can have deleterious effects on polycrystal properties [38]. Hence, the objective of the present research was to understand the atomistic relationship between the local structure and the point-defect energetics at the grain boundary interface in HCP materials, such as Ti, Mg, and Zr. Molecular statics (MS) simulations of Ti, Mg, and Zr bicrystals were used for various $[1\bar{2}10]$ and $[0\bar{1}10]$ tilt grain boundaries to clarify the role of the interface character on point-defect energetics. Of particular interest was how the grain boundary character in HCP materials affects the vacancy binding energies and associated spatial variations in the vicinity of the grain boundary.

This paper is organized as follows. Our simulation methodology is briefly summarized in Section 2. The results and discussion section describe the grain boundary energies and corresponding atomic structure at 0 K, the interface's free volume, the point-defect energetics and the correlation with macroscopic and microscopic degrees of freedom. The simulation results reveal several interesting observations: 1) The grain boundary local arrangements and resulting structural units have a significant influence on the magnitude of vacancy binding energies, and the site-to-site variation within each boundary is substantial. 2) Comparing the vacancy binding energies for each site in different c/a ratio materials shows that the binding energy increases significantly with an increase in c/a ratio. For example, in the $[1\bar{2}10]$ tilt axis, Ti and Zr with c/a=1.5811 had a lower vacancy binding energy than Mg with c/a=1.6299. 3) For all grain boundaries in the three materials examined here, there were atoms lying symmetrically along the grain boundary plane that had vacancy binding energies close to or even higher than the bulk values. Consequently, these grain boundaries may not provide pathways for vacancy diffusion. 4) In most grain boundaries examined here, the vacancy binding energies approached bulk values around 5 Å away from the grain boundary center plane. 5) There is no significant correlation between the vacancy binding energy and the atomic free volume. 6) When the grain boundary energies of all 190 boundaries in all three elements are plotted against the vacancy binding energies of the same boundaries, however, a highly negatively correlation ($r$ = -0.7144) is revealed that has a linear fit with a proportionality constant of -25 Å$^2$.

## II Methodology

To investigate grain boundary sink efficiency in different c/a ratio materials, we employed MS simulations using embedded atom method (EAM) potentials. Initially, a database of 190 minimum energy STGBs of Ti, Mg, and Zr with the tilt axes as $[1\bar{2}10]$ and $[0\bar{1}10]$ was generated with MS simulations, which were performed using the classical molecular dynamics code Large-scale Atomic/Molecular Massively Parallel Simulator (LAMMPS).[39] Here, the analysis cell consisted of a standard bicrystal cell with a single grain boundary that divides the HCP crystal into two single crystals, as shown in Figure 1a. The initial single crystals were created with x, y and z along the $[0\bar{1}10]$, $[0001]$, and $[1\bar{2}10]$ directions, respectively, for the $[1\bar{2}10]$ tilt axis; and along the $[1\bar{2}10]$, $[0001]$, and $[0\bar{1}10]$ directions, respectively, for the $[0\bar{1}10]$ tilt axis. Then, the upper half crystal was rotated clockwise and the lower half crystal counter-clockwise by angle θ with respect to the tilt axis, as shown in Figure 1b. Several successive rigid body translations, followed by an atom-deletion technique and energy minimization using a non-linear conjugate method,[40–42] were used to generate the final relaxed minimum structure with the grain boundary plane along the x-z plane, as shown in Figure 1c. This procedure was replicated to generate several 0 K minimum-energy grain boundary structures for Mg, Ti, and Zr. The EAM potentials of Sun et al.[43] for Mg, Zope and Mishin[44] for Ti, and Mendelev and Auckland[45] for Zr were used.

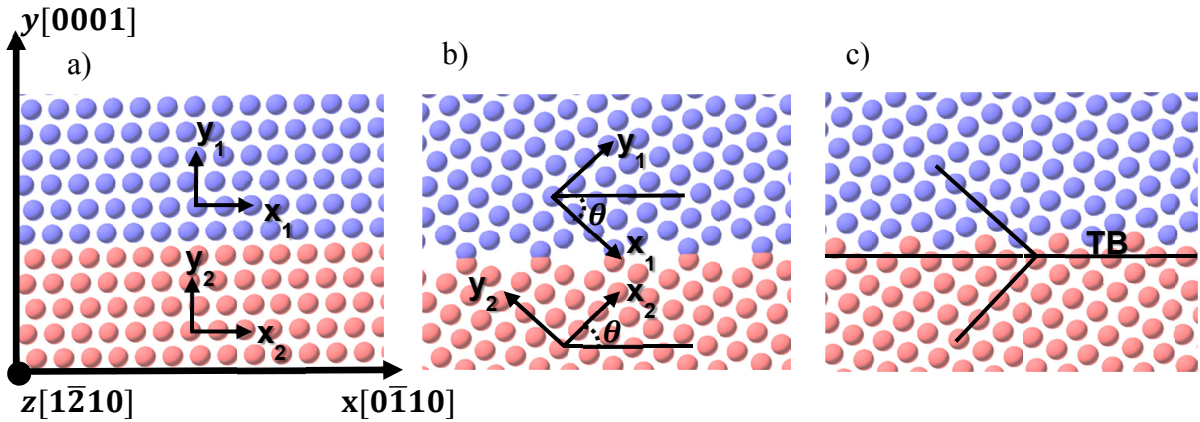

Figure 1: a) Single crystal model with x, y and z along the $[0\bar{1}10]$, $[0001]$, and $[1\bar{2}10]$ directions, respectively; b) Upper half of crystal rotated θ (~42°) clockwise and lower half of crystal rotated counter-clockwise; and c) Final grain boundary structure: rotated crystal is minimized with 1 pN force on each atom using conjugate gradient algorithm. Note: TB notes the twin boundary.

Table 1: Bulk properties of Ti, Mg, and Zr

|    | Lattice parameter $a$ (Å) | c/a ratio | Cohesive energy $E_{coh}$ (eV/atom) | Bulk atomic volume (Å$^3$/atom) | Bulk vacancy formation energy $E_f$ (eV) |
|----|---------------------------|-----------|-------------------------------------|---------------------------------|------------------------------------------|
| Ti | 2.95                      | 1.581     | -4.850                              | 17.577                          | 1.819                                    |
| Mg | 3.20                      | 1.633     | -1.528                              | 22.829                          | 0.868                                    |
| Zr | 3.23                      | 1.581     | -6.013                              | 23.072                          | 1.385                                    |

Following the grain boundary dataset generation, the role of grain boundary character on sink efficiency for vacancies was assessed by calculating the formation energies for vacancies using MS for the generated 190 STGBs in Ti, Zr, and Mg, whereby a vacancy was placed at all sites within in a block of 3x2 nm near the grain boundary center. The database consisted of 125 [1$\bar{2}$10] and 65 [0$\bar{1}$10] STGBs. The vacancy formation energy for a site α is given by:

$$E_f^\alpha = E_{Gb}^\alpha - E_{Gb} + E_{coh}, \quad (1)$$

where $E_{coh}$ is the cohesive energy/atom of a perfect HCP lattice (see Table 1), and $E_{Gb}^\alpha$ and $E_{Gb}$ are the total energies of the grain boundary simulation cell with and without the vacancy, respectively. The cohesive energy for one atom is added to account for the extra atom in the case of the grain boundary simulation cell without the vacancy.

## III Results and Discussion

### A Grain boundary energy and atomic free volume

Understanding the structure and energy of the grain boundary system is crucial for engineering materials intended for advanced applications because grain boundary properties can vary widely (coherent twin versus low-angle versus high-angle grain boundaries). In this study, a range of grain boundary structures and energies that are representative of some of the variation observed in the grain boundary character distribution of polycrystalline as well as nanocrystalline metals was used to investigate the role of grain boundary character on point-defect energetics, specifically the vacancy binding energy in different c/a ratio materials. Figure 2 shows grain boundary energies as a function of the misorientation angle for the [1$\bar{2}$10] and [0$\bar{1}$10] tilt axes in Mg, Ti, and Zr. The trend observed for the grain boundary energy as a function of a misorientation angle is comparable to what has been previously reported for Mg and Ti[34,35] The energy cusps for the [1$\bar{2}$10] system were identified as ($\bar{1}$013)θ = 32.15°, ($\bar{1}$012)θ = 43.31°, ($\bar{1}$011)θ = 62.06°, and ($\bar{2}$021)θ = 75.21° twin boundaries for magnesium, in order of increasing misorientation angle. Similarly, in the case of the [0$\bar{1}$10] tilt axis, the energy cusps were ($\bar{2}$116), ($\bar{2}$114), ($\bar{2}$112), and ($\bar{2}$111) twin boundaries.

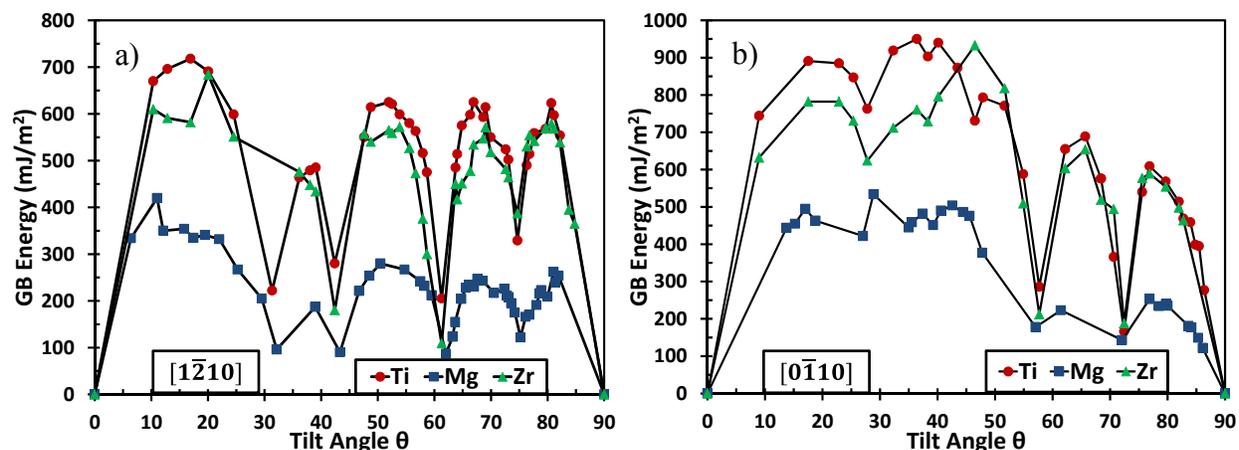

Figure 2: The plot of grain boundary energy as a function of grain boundary misorientation angle for a) the [1$\bar{2}$10] tilt axis, and b) the [0$\bar{1}$10] tilt axis in Ti, Mg, and Zr. Note that the energy cusps for [1$\bar{2}$10] system were identified as ($\bar{1}$013), ($\bar{1}$012), ($\bar{1}$011), and ($\bar{2}$021) twin

boundaries, in order of increasing misorientation angle. Similarly, in the case of [0$\bar{1}$10] tilt axis, the energy cusps were ($\bar{2}$116), ($\bar{2}$114), ($\bar{2}$112), and ($\bar{2}$111) twin boundaries.

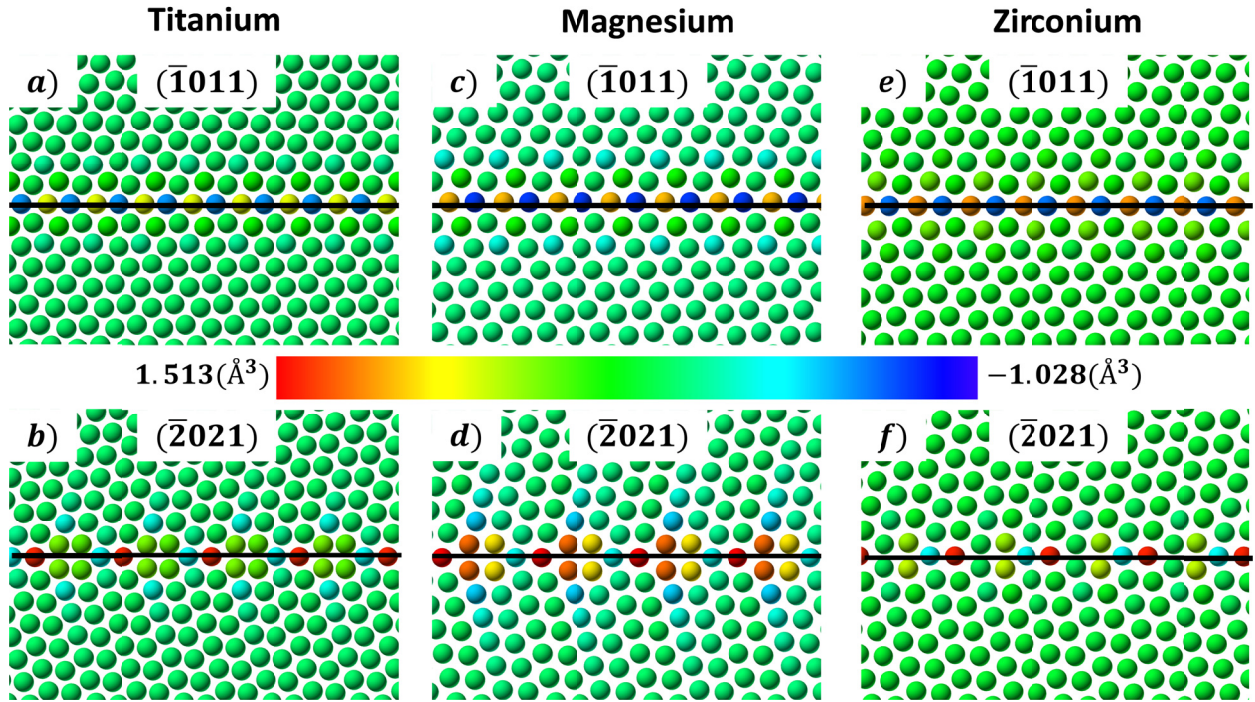

Figure 3: Atomic representation of ($\bar{1}$011) and ($\bar{2}$021) grain boundaries depicting the distribution of the excess Voronoi volume in Ti (a and b), Mg (c and d), and Zr (e and f) for the [1$\bar{2}$10] tilt axis. The bulk Voronoi volume was found to be 17.57, 22.82 and 23.07 Å$^3$ for the Ti, Mg, and Zr, respectively. The black line indicates the twin boundary plane.

The structure-energy correlation can provide more details about the variation in grain boundary energies, as each grain boundary has characteristic SUs describing its atomistic morphology. Low-angle boundaries can be represented by an array of discrete dislocations. However, at higher misorientation angles (high-angle grain boundaries), the dislocation cores overlap, and dislocations rearrange to minimize the boundary energy. The resulting grain boundary structures are often characterized by grain boundary dislocations or SUs.[5,34,35] Figure 3 shows the spatial distribution of atomic excess volume for ($\bar{1}$011) and ($\bar{2}$021) grain boundaries with the [1$\bar{2}$10] tilt axis in Ti, Mg and Zr. Similarly, Figure 4 shows the spatial distribution of excess atomic volume for ($\bar{2}$112) and ($\bar{2}$116) grain boundaries with the [0$\bar{1}$10] tilt axis in Ti, Mg and Zr. Notice that the atoms far away from the boundary are white (0 Å$^3$ excess Voronoi volume), indicating that there is no atomic volume difference over the bulk lattice. The bulk Voronoi volume was found to be 17.57, 22.82, and 23.07 Å$^3$ for the Ti, Mg, and Zr, respectively. The excess Voronoi volume is highest/lowest (tensile versus compressive) at the grain boundary center and converges to the bulk Voronoi volume as distance from the grain boundary increases. Interestingly, the ($\bar{1}$011) and ($\bar{2}$112) plane twin boundaries in Mg and Zr have higher/lowest (tensile versus compressive) excess Voronoi volumes compared to the Ti twin boundary, potentially due to the larger interplanar spacing. That is, near the twin boundary, the interplanar spacing changes and can increase or decrease associated atomic volume due to twinning dislocations (shuffling of atoms).[46,47] In the case of ($\bar{2}$021) and ($\bar{2}$116) plane grain boundaries, Mg has higher/lowest (tensile versus compressive) excess Voronoi volumes compared to Ti and

Zr. In turn, this grain-boundary metric can be correlated to other energetics associated with the grain boundaries to derive a structure-property relationship.

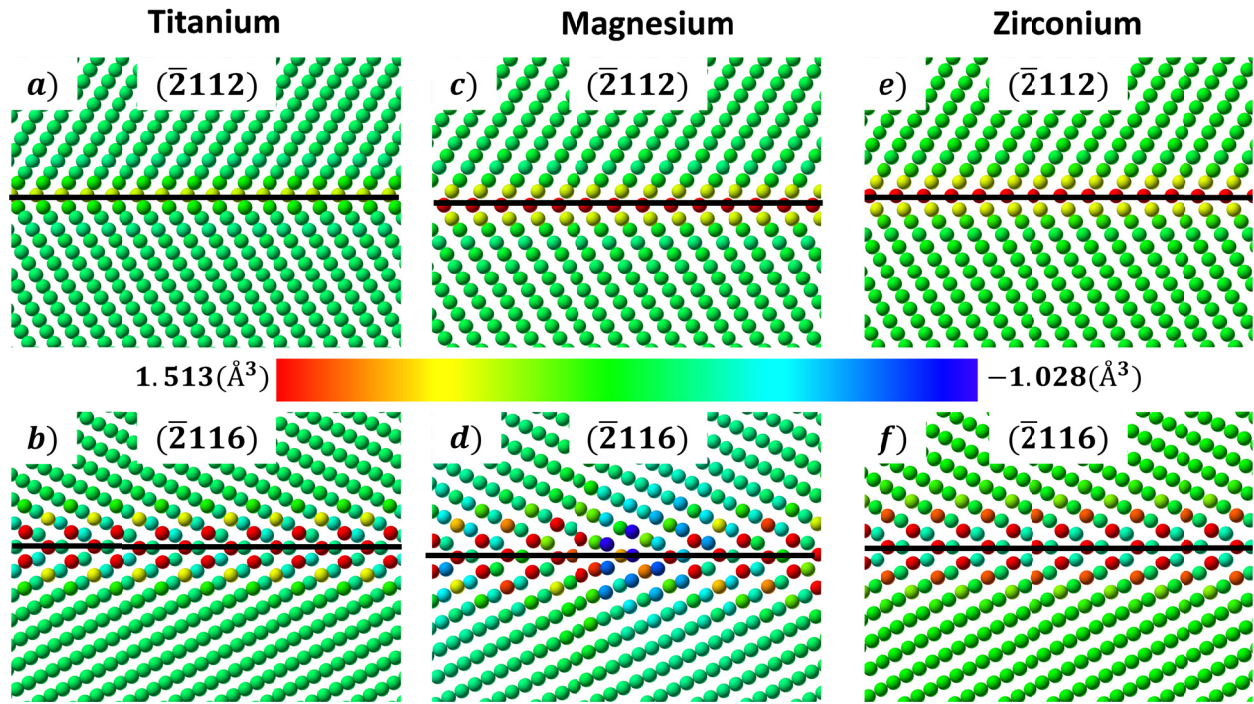

Figure 4: Atomic representation of ($\bar{2}$112) and ($\bar{2}$116) grain boundaries depicting the distribution of the excess Voronoi volume in Ti (a and b), Mg (b and c), and Zr for the [0$\bar{1}$10] tilt axis. The black line indicates the twin boundary plane.

**B Vacancy binding energy**

Molecular statics (MS) was used to examine the vacancy binding energy as a function of the local atomic structure and distance from the grain boundary center. Here, the vacancy was placed at all sites within in a block of 3x2 nm near the grain boundary. Then, the change in the vacancy binding energies with increasing distance from the grain boundary center was used to quantify the nonlocal length scale associated with the vacancy binding. The vacancy binding energy ($E_b$) is essentially the formation energy of a particular site in the grain boundary region, which is normalized with the bulk formation energy ($E_b = E_f - E_f(bulk)$, where $E_f(bulk)$=1.819, 0.868 and 1.385 eV for Ti, Mg, and Zr, respectively. Figures 5-7 show the spatial distribution of vacancy binding energies in selected grain boundaries, which is equivalent to representing the variation of vacancy formation energy for a grain boundary system after removing the bulk contribution. Notice that the atoms far away from the boundary are white (0 eV vacancy binding energy), indicating that there is no energy difference over the bulk lattice. For the ($\bar{1}$011) twin boundary, the minimum vacancy binding energy was found for Ti, Mg, and Zr to be -0.438 eV, -0.157 eV, and -0.365 eV, respectively (see Figures 5 and 7a). Similarly, for the ($\bar{2}$112) plane twin boundary, the minimum vacancy binding energy was found for Ti, Mg, and Zr to be -0.294 eV, -0.167 eV, and -0.134 eV, respectively (see Figure 6). Interestingly, the vacancy binding energy in the ($\bar{1}$011) and ($\bar{2}$112) plane twin boundaries were inversely proportional to the excess free volume. For example, the ($\bar{1}$011) and ($\bar{2}$112) plane twin boundaries in Ti had more negative vacancy binding energies and the lowest free volume when compared with Mg or Zr. In the case of the ($\bar{2}$112) plane twin boundary in all three materials, the vacancy binding energy for the 1$^{st}$ layer was lower than that for 0$^{th}$ layer, indicating that the grain boundary center is not necessarily the sink for vacancy. However, the ($\bar{1}$011) plane twin boundary in all three materials exhibited lower vacancy binding energy for the 0$^{th}$ layer when compared with the 1$^{st}$ layer, suggesting a strong correlation between the grain boundary structural unit and anisotropy associated with the vacancy binding energies. On average, the vacancy binding energies approached bulk values between 3 to 4 layers away from the grain boundary center for the ($\bar{1}$011) and ($\bar{2}$112) plane twin boundaries in all different c/a ratio materials examined here.

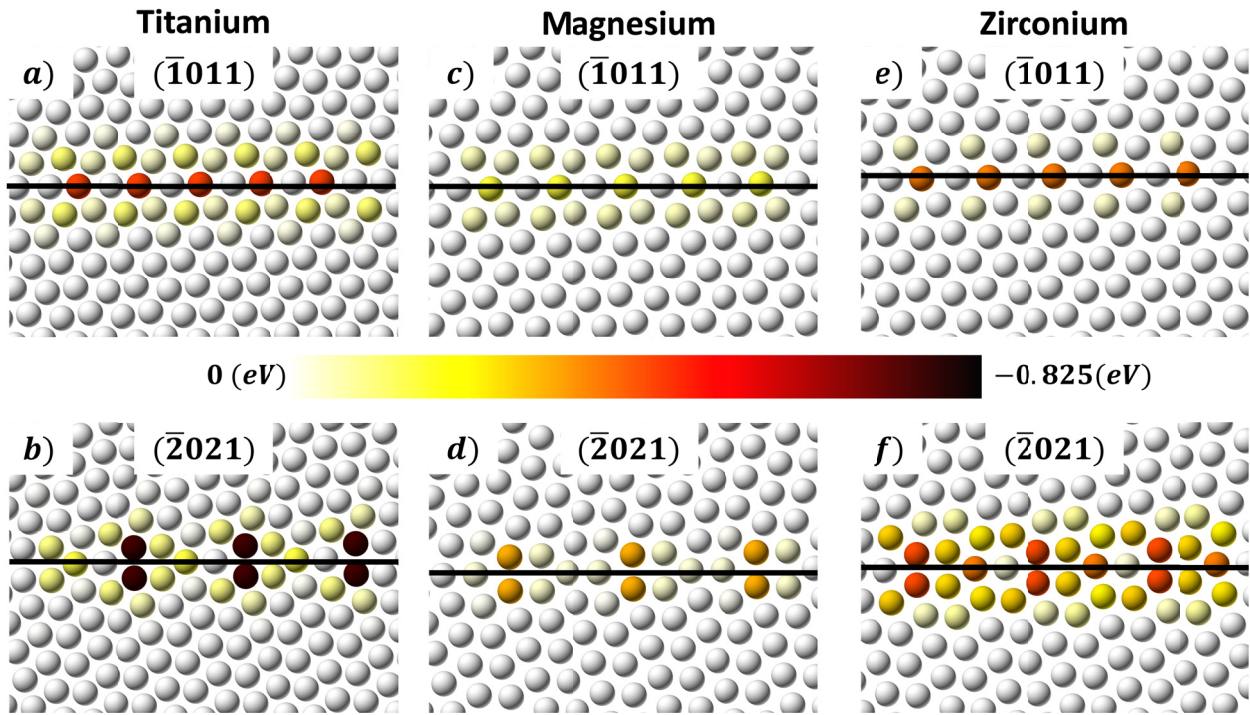

Figure 5: Atomic representation of ($\bar{1}$011) and ($\bar{2}$021) grain boundaries depicting the distribution of vacancy binding energies in Ti (a and b), Mg (c and d), and Zr (e and f) for the [1$\bar{2}$10] tilt axis. The vacancy binding energy in bulk was found to be 1.819, 0.868, and 1.385 eV for Ti, Mg, and Zr, respectively. The black line indicates the twin boundary plane.

For the ($\bar{2}$021) grain boundary, the minimum vacancy binding energy was found for Ti, Mg, and Zr to be -0.743 eV, -0.304 eV, and -0.425 eV, respectively (see Figures 5 and 7b). Similarly, for the ($\bar{2}$116) grain boundary, the minimum vacancy binding energy was found for Ti, Mg, and Zr to be -0.777 eV, -0.825 eV, and -0.435 eV, respectively (see Figure 6). The vacancy binding energy results of ($\bar{2}$021) and ($\bar{2}$116) grain boundaries indicate that there is no significant correlation between the vacancy binding energy and the free volume. For example, the ($\bar{2}$021) grain boundary in Ti and Zr had similar free volume but significantly different vacancy binding energies. Hence, vacancy binding is not correlated with only one atomic volume; rather, contributions from surrounding atoms also play a role in deciding the potential site for vacancy binding.

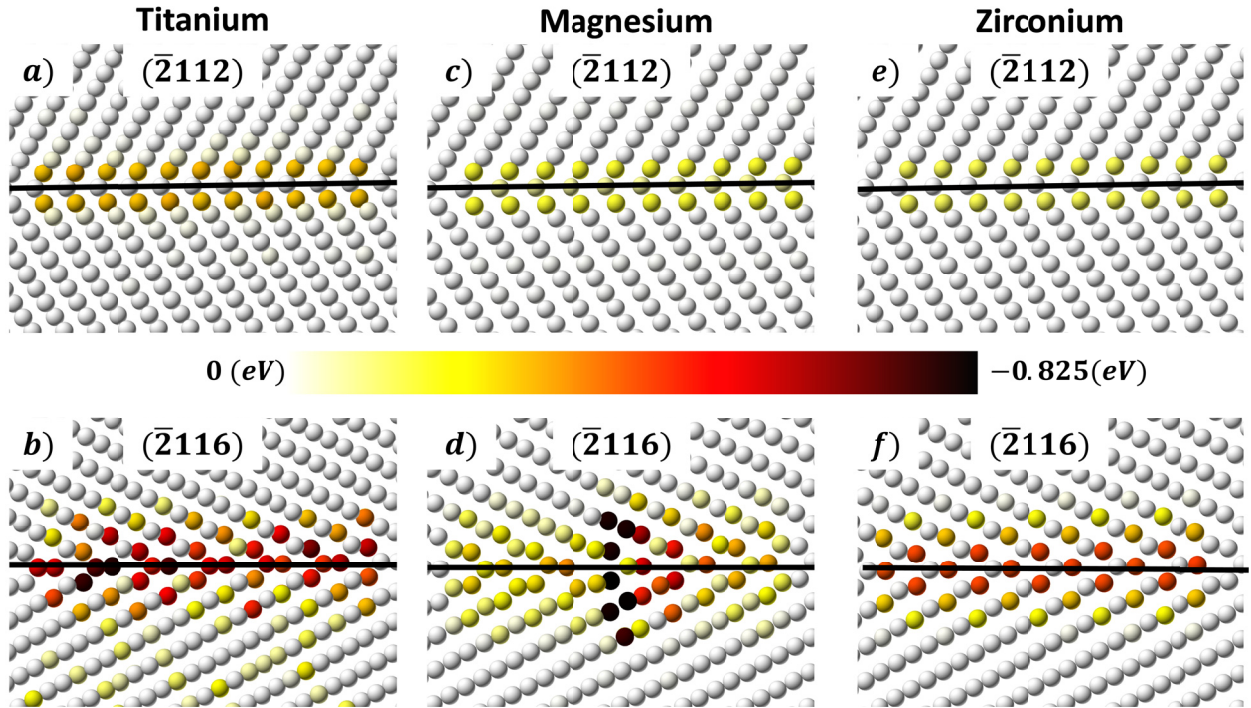

Figure 6: Atomic representation of $(\bar{2}112)$ and $(\bar{2}116)$ grain boundaries depicting the distribution of vacancy binding energies in Ti (a and b), Mg (c and d); and Zr (e and f) for the $[0\bar{1}10]$ tilt axis. Vacancy binding energy in bulk was found to be 1.819, 0.868, and 1.385 eV for Ti, Mg, and Zr, respectively. The black line indicates the boundary plane.

The binding energies of the vacancy can be plotted against the distance from the grain boundary to quantify the evolution of the binding energies near the grain boundary and to quantify the nonlocal length scale associated with the vacancy. Figure 7 is an example of one such plot for vacancy binding energies at various sites for the $(\bar{1}011)$ and $(\bar{2}021)$ grain boundaries. In this plot, the vacancy binding energy was first calculated for each site. Next, a grain boundary region was defined to compare the vacancy binding energies for the $(\bar{1}011)$ and $(\bar{2}021)$ grain boundaries in Ti, Mg, and Zr. Similar to earlier observations, the minimum vacancy binding energy is at $0^{th}$ layer (grain boundary plane) for the $(\bar{1}011)$ grain boundary (Figure 7a) as compared to the $1^{st}$ layer for the $(\bar{2}021)$ grain boundary (Figure 7b) for all c/a ratio materials examined here. Furthermore, in both boundaries, the vacancy binding energies approached bulk values around 5 Å away from the grain boundary center plane.

Overall, for all grain boundaries in the three materials examined here, there were atoms lying symmetrically along the grain boundary plane that had vacancy binding energies close to or even higher than the bulk values, i.e., these grain boundaries may not provide pathways for vacancy diffusion. Finally, these figures show that the local environment strongly influences the vacancy binding energies and that these energies are not independent of one another.

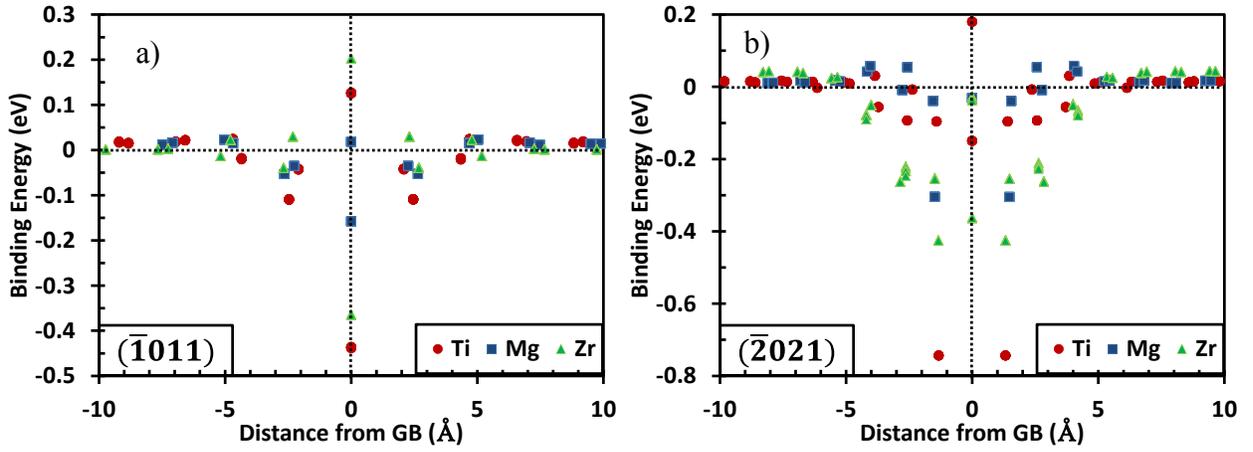

Figure 7: Vacancy binding energy as a function of distance from the grain boundary center: a) the ($\bar{1}$011) plane boundary, where the vacancy binding energy minimum is at the 0$^{th}$ layer of the grain boundary plane, and b) the ($\bar{2}$021) plane boundary, where the vacancy binding energy minimum is at the 1$^{st}$ layer from the grain boundary plane. The vacancy binding energies approached bulk values around 5 Å away from the grain boundary center plane.

## C Correlating grain boundary metrics

The local environment surrounding each atom changes due to interactions with neighboring atoms, which in turn affects the vacancy binding energy and other per-atom properties. In this subsection, we will analyze and correlate calculated vacancy binding energies with grain boundary energies in Ti, Mg, and Zr. The grain boundary energies of all 190 boundaries in all three elements are plotted against the vacancy binding energies of the same boundaries, as shown in Figure 8. The solid line corresponds to a perfect fit with a proportionality constant of -25 Å$^2$. The results indicate that there is a strong correlation between the grain boundary energy and the vacancy binding energy. Furthermore, there is an overall trend of increasing boundary energy with decreasing binding energy, which could be due to atomic scale roughness when two tilted bicrystals face each other. As such, a grain boundary with higher atomic area density will have higher stress fields, which can be relieved through vacancy introduction at the compression site (Voronoi volume of atom having less volume compared to volume of bulk atom). Therefore, grain boundaries with higher grain boundary energies see a significant drop in vacancy binding energy.

Here, the linear correlation coefficient $r$ is used (Eq. 2) to compare the degree of correlation between the binding energy and the grain boundary energy, where $r = 1$ indicates a perfect positive correlation and $r = -1$ indicates a perfect negative correlation. Interestingly, the vacancy binding energy is highly negatively correlated ($r = -0.7144$) with the grain boundary energy.

$$r = \frac{\sum x_i y_i - \frac{\sum x_i \sum y_i}{N}}{\sqrt{\sum x_i^2 - \frac{(\sum x_i)^2}{N}} \sqrt{\sum y_i^2 - \frac{(\sum y_i)^2}{N}}} \qquad (2)$$

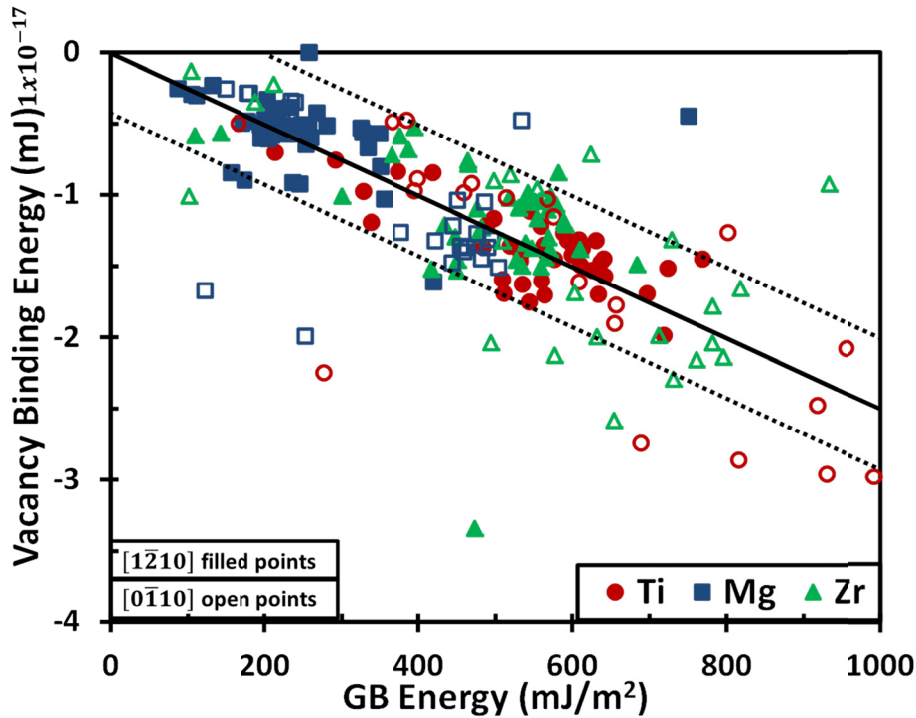

Figure 8: The grain boundary energies of all 190 boundaries in all three elements are plotted against the vacancy binding energies of the same boundaries. Closed data points are for the $[1\bar{2}10]$ tilt axis and open data points for the $[0\bar{1}10]$ tilt axis. Interestingly, the vacancy binding energy is highly negatively correlated ($r = -0.7144$) with the grain boundary energy.

## IV Conclusions

Molecular static simulations of 190 symmetric tilt grain boundaries in HCP metals were used to understand the energetics of vacancy segregation, which is important for designing stable material interfaces for endurance in harsh environments. The simulation results reveal several interesting observations:

1) The grain boundary local arrangements and resulting structural units have a significant influence on the magnitude of vacancy binding energies, and the site-to-site variation within the boundary is substantial (Figures 5-6);

2) Comparing the vacancy binding energies for each site in different c/a ratio materials shows that the binding energy increases significantly with an increase in c/a ratio (Figures 6-7). For example, in the $[1\bar{2}10]$ tilt axis, Ti and Zr with c/a=1.5811 have a lower vacancy binding energy than Mg with c/a=1.6299;

3) For all grain boundaries in the three materials examined here, there were atoms lying symmetrically along the grain boundary plane that had vacancy binding energies close to or even higher than the bulk values, i.e., these Grain boundaries may not provide pathways for vacancy diffusion (Figures 5-7);

4) In most grain boundaries examined here, the vacancy binding energies approached bulk values around 5 Å away from the grain boundary center plane (Figure 7);

5) There is no significant correlation between the vacancy binding energy and the atomic free volume; and

6) When the grain boundary energies of all 190 boundaries in all three elements are plotted against the vacancy binding energies of the same boundaries, a highly negative correlation ($r = -0.7144$) is revealed that has a linear fit with a proportionality constant of -25 Å$^2$ (Figure 8).

In summary, these new atomistic perspectives provide a physical basis for recognizing the incipient role between the grain boundary character and vacancy binding energies in HCP materials. This is significant for applications where extreme environment damage generates lattice defects and grain boundaries act as sinks for both vacancies and interstitial atoms.

**Acknowledgement**

The authors gratefully acknowledge support from the Office of Naval Research and the Air Force Office of Scientific Research under contracts N000141110793 and FA9550-13-1-0144, respectively.